\documentclass[aps,prb,preprint]{revtex4}

\usepackage{graphicx}
\usepackage{dcolumn}
\usepackage{amsmath}
\usepackage{amssymb}
\usepackage{wasysym}
\usepackage{color}

\usepackage{mathtools}

\begin{document}

\title{Electronically driven collapse of the bulk modulus in $\delta$-plutonium
}

\author{N.~Harrison
}

\affiliation{Mail~Stop~E536,~Los~Alamos~National Labs.,Los~Alamos,~NM~ 87545
}
\date{\today}

\begin{abstract}
Plutonium metal exhibits an anomalously large softening of its bulk modulus at elevated temperatures that is made all the more extraordinary by the finding that it occurs irrespective of whether the thermal expansion coefficient is positive, negative or zero --- representing an extreme departure from conventional Gr\"{u}neisen scaling. We show here that the cause of this softening is the compressibility of plutonium's thermally excited electronic configurations, which has thus far not been considered in thermodynamic models. We show that when compressible electronic configurations are thermally activated, they invariably give rise to a softening of bulk modulus regardless of the sign their contribution to the thermal expansion. The electronically driven softening of the bulk modulus is shown to be in good agreement with elastic moduli measurements performed on the gallium-stabilized $\delta$ phase of plutonium over a range of temperatures and compositions, and is shown to grow rapidly at small concentrations of gallium and at high temperatures, where it becomes extremely sensitive to hydrostatic pressure.
\end{abstract}
\maketitle

\section{Introduction}

Plutonium (Pu) has the richest phase diagram amongst the metallic elements, and, as a consequence, has proven to be the most challenging to grasp.\cite{moore2009,hecker2000,smith1983,hecker2004} In recent years, considerable advances have been made towards understanding some of Pu's unusual thermodynamic properties, such as its anonamously enhanced electronic heat capacity for an element\cite{lashley2003,shim2007,zhu2007,lanata2015} and its invar-like negative thermal expansion coefficient.\cite{lawson2002,lawson2006,harrison2019} The softening of the bulk modulus,\cite{soderlind2010,suzuki2011,freibert2012,migliori2016,nadal2010} by contrast, which occurs $\sim$~50\% more rapidly with increasing temperature (relative to the melting temperature) than regular solids,\cite{digilov2019,ida1969,born1939,varshni1970,rose1984,anderson1989,ida1970} has continued to remain a mystery. The increased likelihood that the entirety of the bulk modulus softening cannot be explained by phonons alone has led to the suggestion of an unconventional contribution originating from electronic degrees-of-freedom.\cite{migliori2016,lawson2019}

A natural candidate for electronic degrees-of-freedom in Pu is provided by its unstable $5f$-electron atomic shell, which has been shown to allow Pu to exist in a greater number of near degenerate electronic configurations\cite{eriksson1999,wills2004,svane2007} and oxidation states\cite{windorff2017} than other actinides and rare earths. Direct experimental evidence for the presence of multiple near-degenerate electronic configurations in $\delta$ phase Pu ($\delta$-Pu) has been provided by way of x-ray spectroscopy\cite{booth2012} and neutron scattering experiments.\cite{janoschek2015} While it has been argued on the basis of sophisticated electronic structure models that virtual valence fluctuations cause extensive mixing between these configurations,\cite{savrasov2001,shim2007,zhu2007,lanata2015} a strong thermally-activated component has been indicated by way of thermal expansion experiments\cite{lawson2002,lawson2006} and, more recently, by way of temperature-dependent magnetostriction experiments.\cite{harrison2019} 

%
%

We show here that a crucial factor in causing multiple electronic configurations to contribute significantly to the bulk modulus is their compressibility, which we show to give rise to a previously unknown yet significant electronic contribution to the bulk modulus when excited electronic configurations are thermally activated. Because this contribution is both negative and quadratic in the size of the difference in equilibrium volume (manifesting itself as a partial pressure) between the excited configurations and the lattice, it invariably leads to a softening of the bulk modulus with increasing temperature. Using a form for the free energy recently adapted from measurements of different thermodynamic quantities,\cite{lawson2006,harrison2019} we show that the uncovered electronically driven softening of the bulk modulus is in agreement with temperature-dependent and Ga concentration-dependent resonant ultrasound spectroscopy results, in which Ga is used to stabilize the $\delta$ phase.\cite{suzuki2011,freibert2012,migliori2016,soderlind2010} The softening is shown to become especially large in $\delta$-Pu stabilized with small concentrations of Ga at temperatures well above room temperature, where it is further predicted to undergo a collapse under hydrostatic pressure.

\section{Results} 
\subsection{Origin of the bulk modulus softening}

\begin{figure}[!!!!!!!htbp]
\centering 
\includegraphics*[width=.35\textwidth]{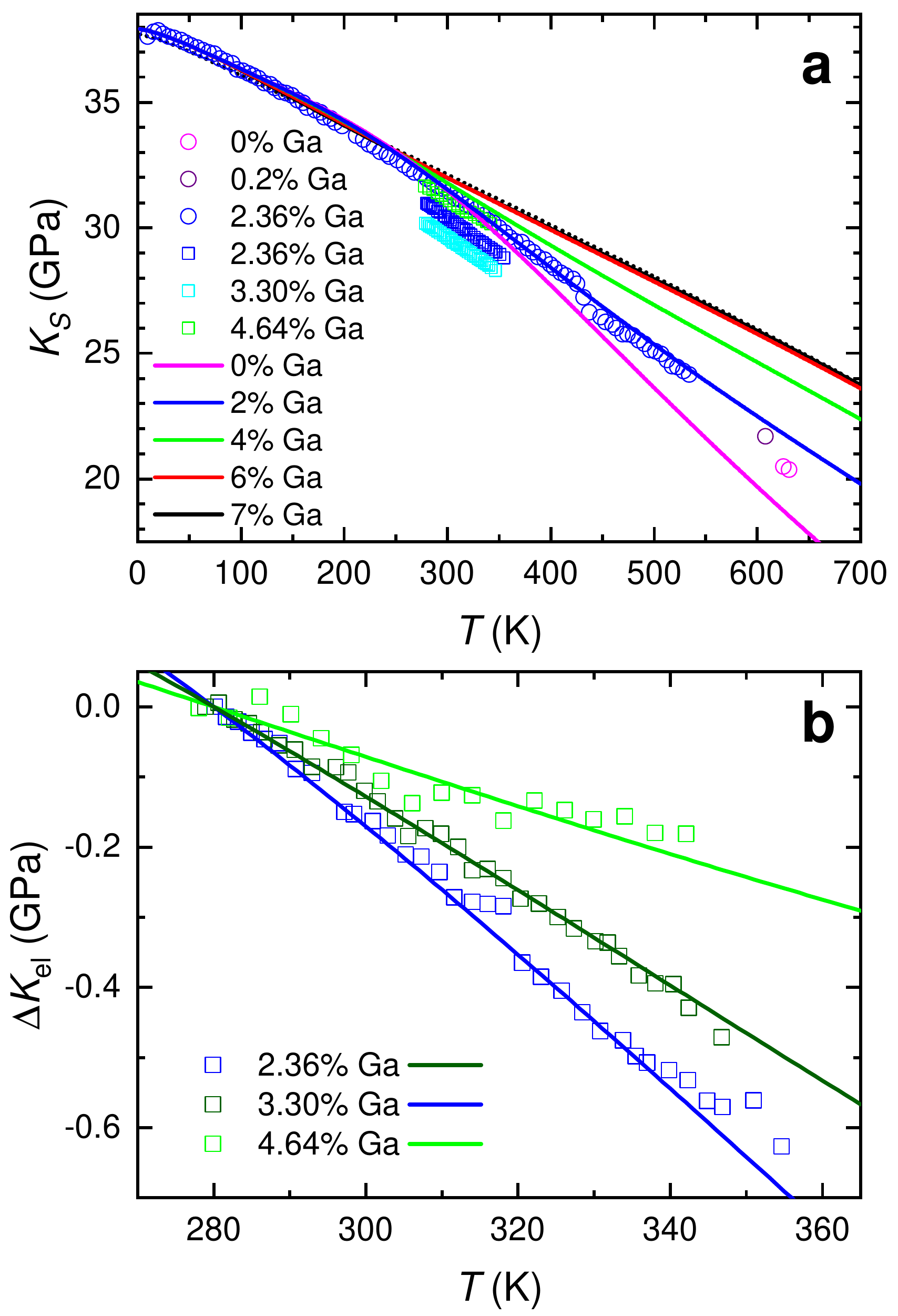}
\caption{{\bf a} A comparison of experimental adiabatic bulk modulus data for different compositions $x$ of $\delta$-Pu$_{1-x}$Ga$_x$ as indicated\cite{soderlind2010,migliori2016,suzuki2011,freibert2012} with the model calculations of $K_S=\gamma K_T$ (where $K_T=K_{\rm el}+K_{\rm ph}$ is the isothermal bulk modulus and $\gamma$ is plotted in the Appendix) for different compositions, as indicated. As a point of reference, the black dotted line is the fitted functional form of $K_{\rm ph}$ (for $b=$~18); we have added this to $K_0=$~37.7~GPa in order to bring it into alignment with the other curves at $T\approx$~10~K. {\bf b} A comparison of the experimentally measured change in temperature-dependent electronic contribution to the bulk moduli $\Delta K_{\rm el}=K_{\rm el}(T)-K_{\rm el}(280~{\rm K})$ for $x=$~2.36\%, 3.30\% and 4.64\% (colored squares), having subtracted the measured values at $T=$~280~K and the calculated change $\Delta K_{\rm ph}=K_{\rm phl}(T)-K_{\rm ph}(280~{\rm K})$ in $K_{\rm ph}$ (again for $b=$~18) relative to that at $T=$~280~K (assuming $K_{\rm ph}$ to be independent of $x$), with the equivalent change $\Delta K_{\rm el}$ in the electronic contribution (colored lines) relative to that at $T=$~280~K calculated using Equation~(\ref{bulkmodulusequation}).  For non-integer values of $x$, calculated values of $K_{\rm el}$ are interpolated in $x$.}
\label{experimentalbulkmod}
\end{figure}

\begin{figure}[!!!!!!!htbp]
\centering 
\includegraphics*[width=.4\textwidth]{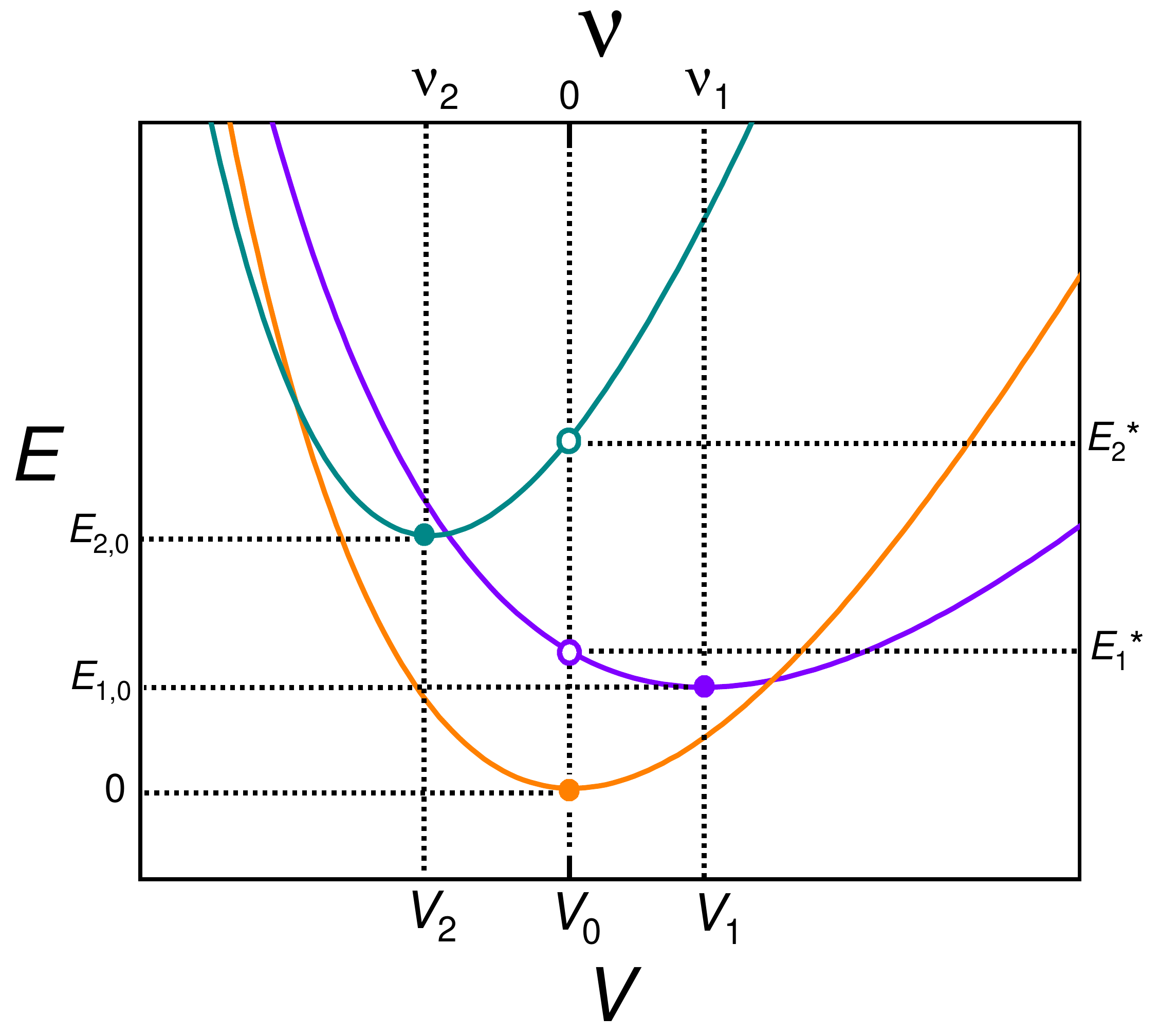}
\caption{Schematic $E(V)$ curves (lower axis) or $E(\nu)$ curves (upper axis) as described in the text according to Equation~(\ref{cohesionvolume}), showing an approximately parabolic form.  The energy minima $E_{i,0}$ and the energies $E^\ast_i$ at $V=V_0$ or $\nu=0$, in relation to the volumes $V_i$ or volume strains $\nu_i$ are also indicated. 
}
\label{schematic}
\end{figure}

Studies of the thermodynamic properties of Pu have shown that their temperature dependences can be modeled by considering a partition function of the form\cite{migliori2016,harrison2019,lawson2006,lawson2002} $Z_{\rm el}=\sum_{i}{\rm e}^{-\frac{E_i}{k_{\rm B}T}}$, where $i$ refers to different electronic configurations with fixed energies $E_i$ and atomic volumes $V_i$. Use of such a partition function for modeling thermodynamic quantities is warranted under circumstances where the mixing between different electronic configurations attributable to valence fluctuations is sufficiently small for the higher energy configurations to be thermally activated.\cite{harrison2019,lawrence1981,wohlleben1984} We find, however, that whereas the consideration of $E_i$ and $V_i$ as fixed and independent quantities is a reasonable approximation for modeling the thermal expansion, heat capacity and the magnetostriction, \cite{harrison2019,lawson2006,lawson2002} this is not the case when considering the bulk modulus (see Fig.~\ref{experimentalbulkmod}). Since the relationship between $E_i$ and $V$ forms the basis of the definition of the bulk modulus of a material, neglect of this relationship has the potential to cause entire terms to be missing from the equation of state. Electronic structure calculations have shown that $E_i$ and $V$ are inextricably linked for each electronic configuration of Pu and other actinides,\cite{svane2007,eriksson1999} 
making $E_i(V)$ a function of $V$, or, equivalently, $E_i(\nu)$ a function of the volume strain $\nu=\frac{V}{V_0}-1$ (see schematic in Fig.~\ref{schematic}). A more generalized form 
\begin{equation}\label{electronictruncpartitionfunction}
Z_{\rm el}(\nu)=\sum_{i}{\rm e}^{-\frac{E_i(\nu)}{k_{\rm B}T}}
\end{equation}
for the partition function that preserves information relating to the compressibility is therefore required.
 
Electronic structure calculations have show that the $E_i(\nu)$ curves of the different configurations (illustrated in Fig.~\ref{schematic}) are approximately parabolic,\cite{svane2007,eriksson1999} enabling them to be represented in the reduced form
\begin{equation}\label{cohesionsmall}
\lim_{\nu-\nu_i\rightarrow0}E_i(\nu)=E_{i,0}+\frac{K_{i}~[\nu-\nu_i]^2}{2N}
\end{equation}\\
for small $\nu$, where $N$ is the atomic density and each of their contributions to the bulk modulus is given by $K_i(\nu)=N\frac{\partial^2E_i(\nu)}{\partial\nu^2}$ evaluated at $\nu=0$. Here, $E_{i,0}$ is the energy minimum and $\nu_i=\frac{V_i}{V_0}-1$ is the volume strain at the energy minimum for each configuration (see Fig.~\ref{schematic}). For the lowest energy (ground state) configuration, by definition $i=0$, $V_i=V_0$, $\nu_{i}=0$ and $E_{i,0}=0$. The excited electronic configurations, by contrast, are in nonequilibrium states, causing them each to exert a statistical partial pressure $P_i(\nu)=-N\frac{\partial E_i(\nu)}{\partial\nu}$ on the surrounding lattice. In the limit of small strain, this pressure is simply 
\begin{equation}\label{effectivepressure}
\lim_{\nu-\nu_i\rightarrow0}P_i(\nu)=-{K_{i}~[\nu-\nu_i]}.
\end{equation}
The ensemble average of the volume changes associated with these partial pressures is what ultimately drives the positive and negative electronic contributions to the thermal expansion (see Appendix).\cite{harrison2019}

We proceed to calculate the electronic contribution to the bulk modulus by taking the second derivative $K_{\rm el}=\frac{\partial^2F_{\rm el}}{\partial\nu^2}\big|_T$ of the electronic component $F_{\rm el}=-k_{\rm B}T\ln Z_{\rm el}(\nu)$ of the free energy with respect to the volume strain. Using the volume-dependent partition function given by Equation~(\ref{electronictruncpartitionfunction}) and $E_i(\nu)$ curves, we obtain
\begin{widetext}
\begin{equation}\label{bulkmodulusequation}
K_{\rm el}={{\sum_{i}p_i(\nu)K_i(\nu)}}-\frac{\sum_{i}p_i(\nu)P^2_i(\nu)}{Nk_{\rm B}T}
+\frac{\big[\sum_{i}p_i(\nu)P_i(\nu)\big]^2}{Nk_{\rm B}T},
\end{equation}
\end{widetext}
where
\begin{equation}\label{probability}
p_i(\nu)=Z_{\rm el}^{-1}{\rm e}^{-\frac{E_i(\nu)}{k_{\rm B}T}}
\end{equation}
is the probability of occupancy for each configuration.

The first term on the right-hand-side of Equation~(\ref{bulkmodulusequation}) is the probability-weighted sum of bulk moduli that has been assumed in prior models of the multiple electronic configurations in $\delta$-Pu.\cite{lawson2002,lawson2006,migliori2016} While large changes in $K_i(\nu)$ with $i$, (we discuss the extreme case where $K_i=0$ for an excited configuration in the Appendix)\cite{lawson2006,lawson2019} have the potential to yield significant changes in $K_{\rm el}$ with temperature, the bulk moduli contributions of all of the electronic configurations obtained by density functional theory are found to all be very similar at $\nu=0$.\cite{eriksson1999,svane2007} In Table~\ref{table2}, we find these to have a mean value of $\bar{K}_i(\nu=0)=$~28.2~GPa and a standard deviation of only $\sigma K_i(\nu=0)=$~5.0~GPa.
The first term in Equation~(\ref{bulkmodulusequation}) is therefore not expected to lead to significant changes of the bulk modulus with increasing temperature.


The second term on the right-hand-side of Equation~(\ref{bulkmodulusequation}) has the potential to lead to much larger changes in the bulk modulus of  $\delta$-Pu with increasing temperature, making it the primary motivation of the present study. The origin of this term is the statistical partial pressure $P_i(\nu)$ between thermally excited configurations and the ground state that occurs as a result of their equilibrium volume strains $\nu_i$ being nonzero. For sufficiently small total strains $\nu-\nu_i$, this partial pressure is linear as shown in Equation~(\ref{effectivepressure}). Because the partial pressure produces a negative quadratic contribution to the electronic bulk modulus in Equation~(\ref{bulkmodulusequation}), it implies that thermally fluctuating electronic configurations invariably lead to a softening of the bulk modulus irrespective of whether $\nu_i>0$, as for a positive contribution to the thermal expansion, or $\nu_i<0$, as for a negative contribution to the thermal expansion (see Appendix). 

The third term on the right-hand-side of Equation~(\ref{bulkmodulusequation}) is also determined by $P_i(\nu)$. However, because the probability factors $p_i(\nu)$ in this term are multiplied together, its overall contribution to the bulk modulus is weaker than that of the second term.

For the effect of hydrostatic pressure on the bulk modulus, this we estimate by taking the third derivative of the free energy with respect to $\nu$ and considering $\frac{\partial}{\partial P}=-K_T^{-1}\frac{\partial}{\partial\nu}$, where $K_T$ is the isothermal bulk modulus, whereupon we obtain
\begin{widetext}
\begin{equation}\label{pressurederivative}
K^\prime=\frac{\partial K_{\rm el}}{\partial P}\bigg|_T\approx\frac{1}{K_T[Nk_{\rm B}T]^2}\bigg[{{{\sum_{i}p_i(\nu)P_i^3(\nu)}}}-3{\big[\sum_{i}p_i(\nu)P^2_i(\nu)\big]\big[\sum_{i}p_i(\nu)P_i(\nu)\big]}
+2{\big[\sum_{i}p_i(\nu)P_i(\nu)\big]^3}+\delta\bigg]
\end{equation}
\end{widetext}
--- once again assuming the parabolic approximation (see Appendix). The first term on the right-hand-side in Equation~(\ref{pressurederivative}), which originates from the derivative of the anomalous softening (i.e. the second) term in Equation~(\ref{bulkmodulusequation}), is found to dominate over the other terms. Its dominance implies that the sign and magnitude of the change in bulk modulus under pressure is determined almost entirely by the partial pressures of the electronic configurations, which in turn depend on the signs of $\nu_i$. The cubic dependence on $P_i(\nu)$ implies $K^\prime$ has a more extreme sensitivity to composition than $K_T$. The last term on the right-hand-side of Equation~(\ref{pressurederivative}), $\delta$, is a correction term (see Appendix) that vanishes in the limit where the bulk moduli of the electronic configurations are the same. 

\subsection{Electronically driven softening estimates}

\begin{figure}
\centering 
\includegraphics*[width=.35\textwidth]{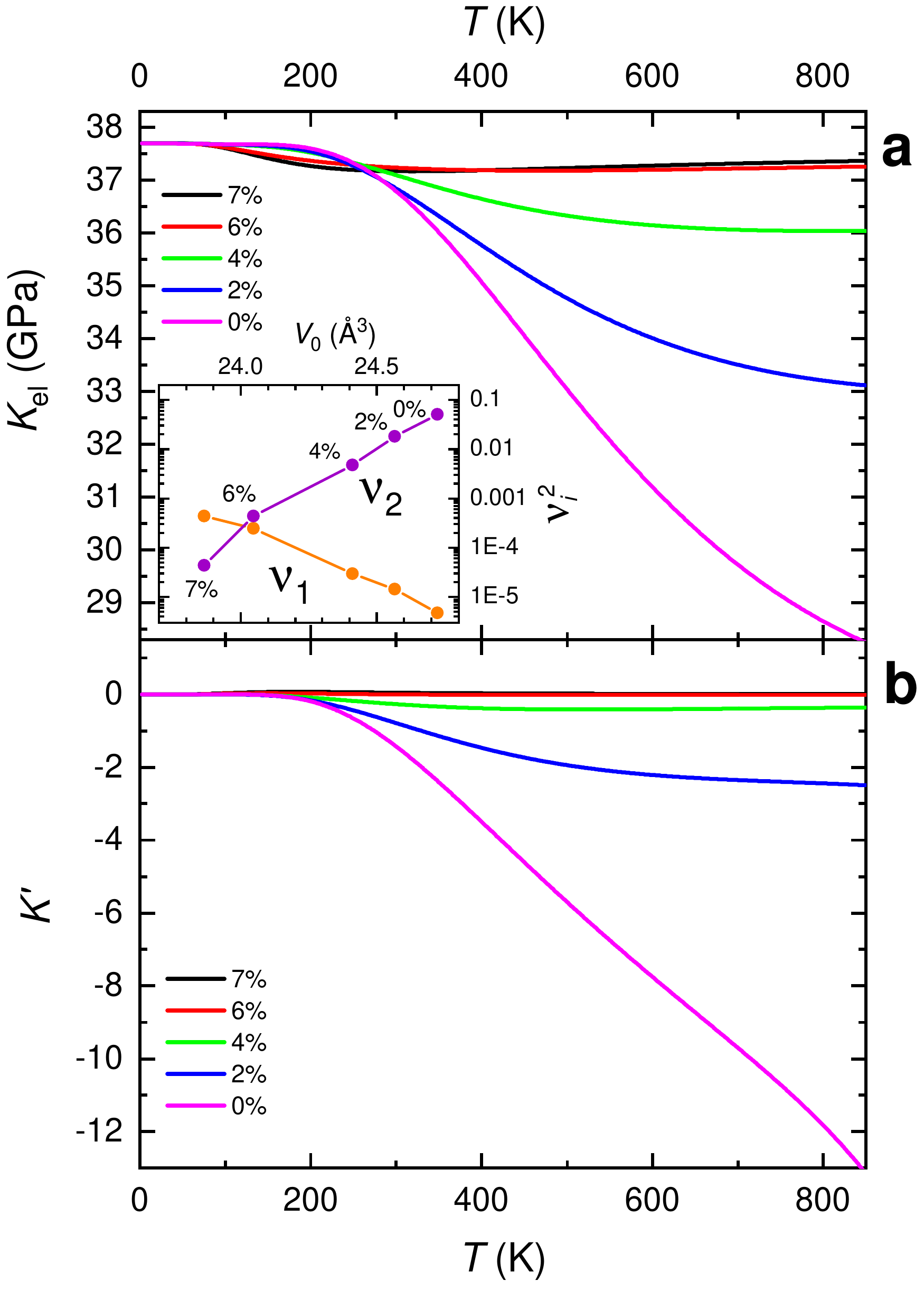}
\caption{{\bf a}, $K_{\rm el}$. of Pu$_{1-x}$Ga$_x$ as a function of temperature $T$ and Ga composition $x$ (colored lines) calculated by way of Equations~(\ref{cohesionsmall}) though (\ref{probability}), using the parameters listed in Table~\ref{table1} for $x=$~2\% and $x=$~7\%. Parameters are extrapolated for $x=$~0\% and interpolated for $x=$~4\% and 6\%. The inset shows $\nu_i^2$ for $i=1$ and 2 versus $V_0$, with the percentage of Ga ($x$) indicated for each point. It is important to note that because $P_i=0$ for the ground state configuration, the denominators of the second and third terms in Equation~(\ref{bulkmodulusequation}) do not lead to a divergence at $T=0$. {\bf b}, Calculated pressure derivative of the bulk modulus $K^\prime$ according to Equation~(\ref{pressurederivative}).}
\label{electronicbulkmod}
\end{figure}

We proceed to estimate the electronic contribution to the bulk modulus and its pressure derivative in Fig.~\ref{electronicbulkmod} from the multiple electronic configurations, by defining $E_i^\ast\approx E_{i,0}+K_i\nu_i^2/2N$ according to Equation~(\ref{cohesionsmall}) and using the approximation $P_i\approx K_i\nu_i$ according to Equation~(\ref{effectivepressure}).\cite{harrison2019} 
Low temperature specific heat measurements have shown that the Debye temperature $\Theta_{\rm D}\approx$~100~K remains largely unchanged as a function of the Ga concentration $x$ used to stabilize the $\delta$ phase,\cite{harrison2019} which is consistent with the parabolic approximation given by Equation~(\ref{cohesionsmall}). We therefore assume that the bulk modulus of the ground state electronic configuration also remains unchanged, and adopt the value $K_0=$~37.7~GPa found in  $\delta$-Pu$_{1-x}$Ga$_x$ for $x=$~2.36\%\cite{suzuki2011} by way of resonant ultrasound measurements. Since the bulk moduli of the excited electronic configurations in $\delta$-Pu are unknown, yet are predicted to fall within a narrow range of possible values in Table~\ref{table2}, we further assume the excited configurations to have bulk moduli $K_i$ that are similar to that of the ground state configuration (see Table~\ref{table1}).

\begin{table}[ht]
\centering
\begin{tabular}{c c c c c}\\
\hline\hline
Quantity & $x=$~2\% Ga &~~~~all $x$~~~~& $x=$~7\% Ga & Units \\ [0.5ex] 
\hline

$K_0=K_1=K_2$&&37.7&&GPa  \\
$\nu_0$&&0&&- \\
$\nu_1$&3.7~$\times$~10$^{-3}$&&2.1~$\times$~10$^{-2}$&- \\
$\nu_2$&-0.13(-0.17)&&-6.7~$\times$~10$^{-3}$&- \\
$E_0^\ast$&&0&& meV \\
$E_1^\ast$&22.8&&41.0& meV \\
$E_2^\ast$&125(121)&&70& meV \\ 
$b$&&18&&-\\
$T_0$&&1.39~$\times$~10$^5$&&K\\
$V_0$&24.57&&23.87&\AA$^3$\\[1ex]
\hline
\end{tabular}
\caption{{\bf Parameters}. Values of parameters used in the calculation of the electronic contribution to the bulk modulus using Equations~(\ref{cohesionsmall}) though (\ref{probability}), respectively. $K_0$ is taken from Suzuki~{\it et al.},\cite{suzuki2011} $\nu_i$ and $E_i^\ast$ values are taken from Harrison~{\it et al.},\cite{harrison2019} $T_0$ values are taken from Lawson\cite{lawson2019} while $V_0$ is taken from diffraction experiments.\cite{lawson2002} For comparison, values of $\nu_2$ and $E_2^\ast$ obtained from fits to the invar model are shown in parentheses for $x=$~2\%.\cite{lawson2006}}
\vspace{2cm}
\label{table1}
\end{table}

In calculating the electronic contribution to the bulk modulus, we use excitation energies $E_i^\ast$ and equilibrium volume strains $\nu_i$ determined\cite{harrison2019} (values listed in Table~\ref{table1}) from fitting thermal expansion\cite{lawson2002,lawson2006} and temperature-dependent magnetostriction\cite{harrison2019} measurements, the results of which were further validated using heat capacity measurements.\cite{harrison2019,lashley2003} The resulting calculations of the electronic contribution to the bulk modulus as a function of $T$ and $x$ using Equations~(\ref{cohesionsmall}) though (\ref{probability}) are shown in Fig.~\ref{electronicbulkmod}a. Both of the previously determined\cite{harrison2019} excited electronic configurations ($E_1^\ast$ and $E_2^\ast$) are found to lead to discernible reductions in the bulk modulus with increasing temperature. Since $E_1^\ast$ leads to a positive thermal expansion\cite{harrison2019} whereas $E_2^\ast$ leads to a negative thermal expansion,\cite{lawson2002,lawson2006} yet both lead to a softening of the bulk modulus, it is the combination of both of these terms that is responsible for the departures from simple Gr\"{u}neisen scaling in $\delta$-Pu.

The degree of bulk modulus softening with temperature is significantly more pronounced for the higher excitation energy $E_2^\ast=$~125~meV (for $x=$~2\%) at small Ga concentrations due to the large ($\approx$~13\%) difference between its equilibrium volume $V_2$ and the ground state volume $V_0$. Very similar results at high temperatures would therefore be obtained on neglecting $E_1^\ast$ and $\nu_1$ and considering only the excitation energy $E_2^\ast$ and volume strain $\nu_2$ determined from the invar model fits.\cite{lawson2002,lawson2006} (see Table~\ref{table1}). According to our calculation, this dominant excitation is predicted to yield a reduction in the bulk modulus that is as large as $\approx$~8~GPa in pure $\delta$-Pu at 700~K; beyond this temperature the $\delta$ phase becomes unstable.\cite{hecker2000}

Turning to the effect of hydrostatic pressure on the bulk modulus, since the leading term on the right-hand-side of Equation~(\ref{pressurederivative}) varies as the cube of the partial pressure, $K^\prime$ is found to be strongly dependent on $x$. The invar contribution, which is characterized by a negative partial pressure, clearly dominates, leading to the prediction of a dramatic collapse of the bulk modulus under pressure and at high temperatures for small concentrations of Ga in Fig.~\ref{electronicbulkmod}b. 


\subsection{Comparison with experiment}

In order to compare the calculations against experimental data, we must also include the phonon contribution to the bulk modulus, which is known to reduce the bulk modulus of most materials by $\approx$~20\% upon reaching $T=T_{\rm m}/2$,\cite{digilov2019,ida1969,born1939,varshni1970,rose1984,anderson1989,ida1970} where $T_{\rm m}$ is the melting temperature ($T_{\rm m}\approx$~912~K in Pu). While a universal model able to accurately describe the reduction in bulk modulus $K_{\rm ph}$  attributable to phonons in all materials has yet to be developed,\cite{digilov2019,ida1969,born1939,varshni1970,rose1984,anderson1989,ida1970} the model of Ida\cite{ida1969,ida1970} has been shown to provide a good description of the heat capacity of $\delta$-Pu at temperatures above room temperature\cite{lawson2019} -- most notably an observed upturn in the heat capacity above $\sim$~600~K. Since the electronic and phonon contributions to the free energy are additive, this should, to a first approximation, be similarly true for derivatives, in which case $K_T=K_{\rm el}+K_{\rm ph}$ (see Appendix) for the isothermal bulk modulus. To compare with the adiabatic bulk modulus $K_S$ obtained by ultrasound measurements, we use the fact that $K_S=\gamma K_T$, where $\gamma\approx1$ for $\delta$-Pu (see Appendix).

In comparing the calculation with the experimental bulk modulus data, only the phonon scaling coefficient $b$ is adjusted. The remainder of the parameters are taken from published results (tabulated in Table~\ref{table1}). Figure~\ref{experimentalbulkmod}a shows that on combining the electronic and lattice vibration contributions, a phonon coefficient $b=$~18 (see Table~\ref{table1} and Appendix) yields a $K_S$ that closely follows the temperature-dependence of bulk modulus measured in $\delta$-Pu$_{1-x}$Ga$_x$ (with $x=$~2.36\%)\cite{suzuki2011} over a broad range of temperatures. Establishing further confidence in the model is the finding that the value of the phonon coefficient $b=$~18 that best fits the phonon part of the bulk modulus is very similar to that $b=$~16 that best fits the anharmonic phonon contribution to the high temperature heat capacity.\cite{lawson2019} For this value of $b$, $K_{\rm ph}$ also accounts for an $\approx$~20\% reduction in $K_S$ with temperature at $T_{\rm  m}/2\approx$~456~K (the remainder coming from the reduction in $K_{\rm el}$), therefore making the phonon contribution to the softening comparable to that in other materials.\cite{digilov2019,ida1969,born1939,varshni1970,rose1984,anderson1989,ida1970}

Having established the approximate form of the (assumed) $x$-independent phonon contribution $K_{\rm ph}$ to the bulk modulus softening, we can proceed to subtract this contribution from measurements of $K_S$ in other samples so as to isolate the electronic contribution $K_{\rm el}$ to the bulk modulus softening, and to investigate its changes with temperature and composition $x$. Figure~\ref{experimentalbulkmod}b shows the $T$-dependence of the residual electronic contribution $K_{\rm el}$ for samples of three different compositions,\cite{soderlind2010} $x=$~2.36\%, 3.30\% and 4.64\%, after having subtracted $K_{\rm ph}$ as well as an offset (see Appendix) to bring the measured curves into alignment at $T=$~280~K. Despite the limited range in temperature of these measurements, significant differences in the temperature dependences are clearly discernible. On comparing the model predictions of $K_{\rm el}$ calculated for the same $x$ compositions using Equation~(\ref{bulkmodulusequation}) with the experimental curves in Fig.~\ref{experimentalbulkmod}b, we find them to be in excellent agreement --- both with regards to the temperature-dependence and the $x$-dependence of the temperature-dependence of the experimental data. Apart from a subtraction of the values of $K_{\rm el}$ at $T=$~280~K that are necessary to eliminate offsets between the experimental curves, no adjustment has been made to $K_{\rm el}$ calculated using Equation~(\ref{bulkmodulusequation}) --- the parameters used are those determined elsewhere by fitting other thermodynamic quantities\cite{harrison2019} (tabulated in Table~\ref{table1}). 

The other compositions, $x=$~0.2\% and $x=$~0\%\cite{freibert2012,migliori2016} in Fig.~\ref{experimentalbulkmod}a, exhibit trends relative to other compositions that are consistent with the model calculations of $K_S$. However, the lack of temperature-dependent data for these compositions means we cannot isolate the electronic component for these compositions. 

\begin{figure}
\centering 
\includegraphics*[width=.35\textwidth]{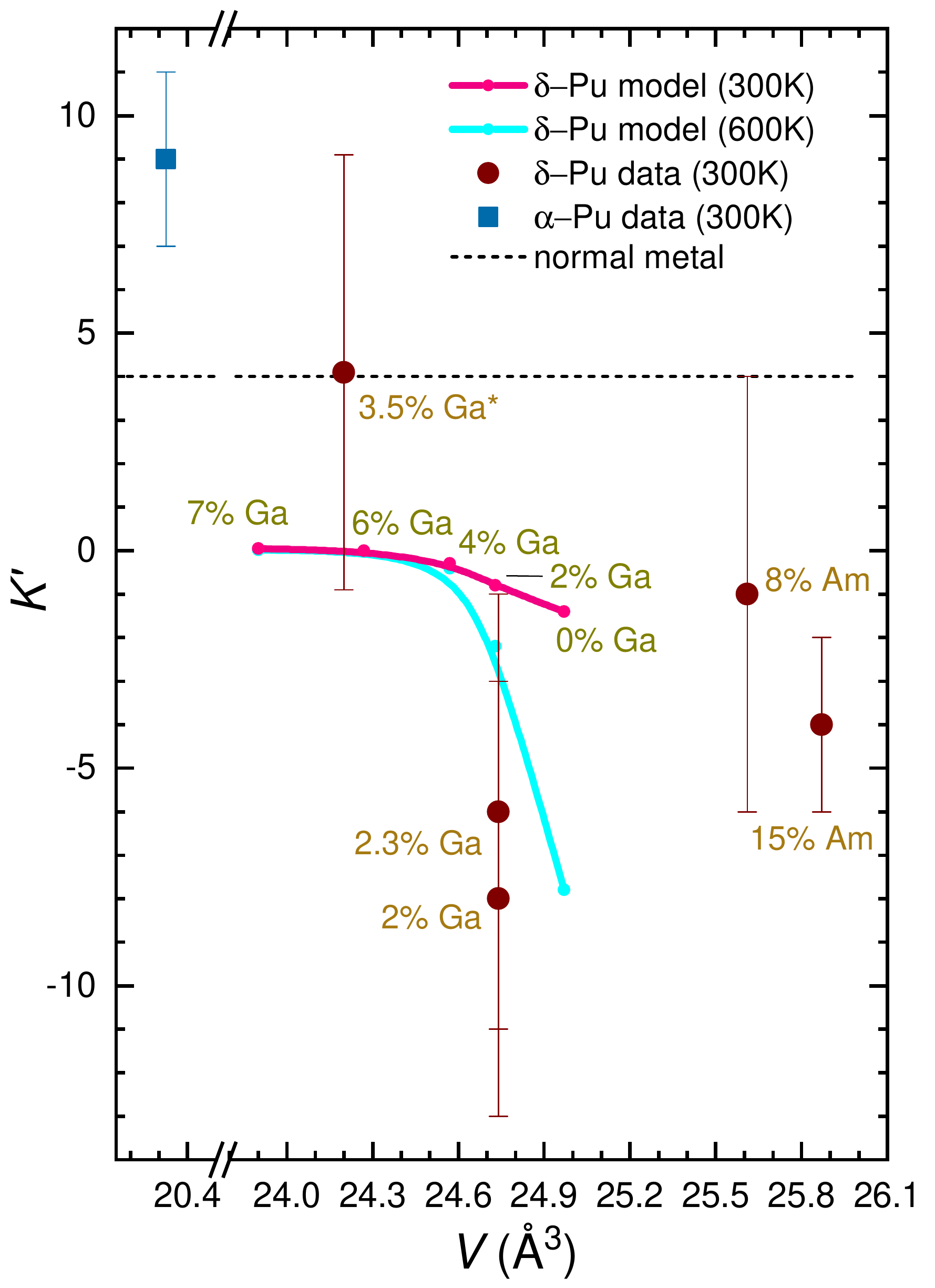}
\caption{Calculated pressure derivative of the bulk modulus $K^\prime$ of Ga-stabilized $\delta$-Pu at 300~K  (pink line and points) according to Equation~(\ref{pressurederivative}) versus atomic volume $V$ at 300~K for different concentrations $x$ of Ga as indicated. This is compared with experimental data for Ga and Am stabilized $\delta$-Pu for different concentrations $x$ as indicated (brown circles). Also shown, for comparison, are the value (light blue square) for $\alpha$-Pu (2.3\% Ga), the calculated Ga-stabilized values $\delta$-Pu values at 600~K (cyan line and points) and the value expected for a normal metal (black dashed line). All of the experiments were performed on $^{239}$Pu with the exception of the sample with $x=$~3.5\%, for which $^{242}$Pu was measured.}
\label{pressurebulkmod}
\end{figure}

Turning once again to the effect of pressure, a fundamental question concerns whether prior measurements of a negative thermal expansion serve as a reliable predictor of a pressure-induced bulk modulus softening.\cite{lawson2006,klosek2008} Existing experimental studies have focused only on the effect of hydrostatic pressure on the bulk modulus at ambient temperature ($\approx$~300~K),\cite{klosek2008,faure2005,faure2010,zhang2019}  the results of which are compared with the model calculations at 300~K (from Fig.~\ref{experimentalbulkmod}b) in Fig.~\ref{pressurebulkmod}. Importantly, Equation~(\ref{pressurederivative}) is found to yield a pressure-induced change in $\delta$-Pu whose negative sign agrees with the softening obtained experimentally; on taking an average of the $\delta$-Pu data points in Fig.~\ref{pressurebulkmod} (including both Ga-stabilized and Am-stabilized $\delta$-Pu), we obtain $K^\prime=$~$-3$~$\pm$~$2$. The calculated value is clearly different from the positive (i.e. stiffening) $K^\prime=+9$~$\pm$~2 value measured in $\alpha$-Pu (plotted for comparison),\cite{faure2010} and the stiffening $K^\prime=+4$ expected for a normal metal. However, the degree of scatter in the experimental data, the magnitude of the error bars and the lack of temperature-dependent data prevent firm conclusions from being reached concerning the absolute magnitude, the doping-dependence or the volume-dependence of $K^\prime$ (see Appendix).

Figures~\ref{electronicbulkmod}b and \ref{pressurebulkmod} show that a significant increase in the observable pressure-induced softening of Ga-stabilized $\delta$-Pu could in principle be achieved by increasing the experimental temperature, which would also enable a more robust verification of the role of excited electronic configurations to be made. Access to higher temperatures would also enable such experiments to be carried out on pure $\delta$-Pu.



\section{Discussion}

By taking into consideration the compressible nature of the previously isolated multiple electronic configurations in $\delta$-Pu,\cite{harrison2019,lawson2002,lawson2006} we have discovered a previously unknown yet significant electronic contribution to the bulk modulus. We find this contribution to be primarily responsible for the excess softening of the bulk modulus in Pu,  the dominant energy scale of which is the same as that previously associated with the invar effect.\cite{lawson2002,lawson2006} The effect of the electronically driven softening is discernible in Ga composition-dependent experiments close to room temperature, but is shown to be strongly enhanced at temperatures substantially above room temperature in $\delta$-Pu$_{1-x}$Ga$_x$ samples with low concentrations of Ga. The bulk modulus is further shown to soften under pressure, as found experimentally,\cite{klosek2008,faure2005,faure2010} and is further predicted to undergo a collapse at low concentrations of Ga and high temperatures. Conversely, the electronic contribution to the softening is expected to be much smaller for samples with large concentrations of Ga at high temperatures, providing an opportunity for the phonon contribution $K_{\rm ph}$ to be more accurately isolated in future studies. 

What constitutes a significant advance in the present approach to modeling the bulk modulus is that the softening is based entirely on the same free energy that has been shown to accurately describe other thermodynamic quantities as a function of temperature and Ga composition. These include the experimentally observed thermal expansion,\cite{lawson2002,lawson2006} magnetostriction\cite{harrison2019} and heat capacity.\cite{lashley2003,lawson2019,harrison2019} It also involves the same energy scale associated with the invar effect and detected in neutron scattering experiments.\cite{janoschek2015} Our results reveal the central role played by statistical thermodynamics in the equation of state in plutonium near ambient pressure.

\section{Appendix}

\subsection{Breakdown of conventional Gr\"{u}neisen scaling}

Whereas in ordinary solids the combination of the volume thermal expansivity $\alpha_v$, the molar heat capacity $C_v$, the isothermal bulk modulus $K_T$ and the molar volume $V_{\rm m}$ equates to a dimensionless Gr\"{u}neisen coefficient\cite{mitra1957,mitra1986}
\begin{equation}\label{gruneisen}
\Gamma=\frac{\alpha_vV_{\rm m}K_T}{C_v},
\end{equation}
simple scaling breaks down in Pu owing to multiple sources of entropy.\cite{lawson2002,lawson2006,harrison2019} 
This is demonstrated spectacularly by the bulk modulus continuing to fall unabated with increasing temperature\cite{suzuki2011,freibert2012,migliori2016,soderlind2010} regardless of whether the thermal expansion coefficient is positive, as is the case for most solids and phases of Pu,\cite{lawson2002,lawson2006} or whether it changes to negative, as occurs for the $\delta$ phase of Pu ($\delta$-Pu) at elevated temperatures and low concentrations of doped gallium (Ga).\cite{lawson2002,lawson2006}

\subsection{Energy-versus-volume}

The calculation of the electronic contribution to the bulk modulus from the second derivative of the free energy requires knowledge of the functional forms of $E_i(\nu)$, for which we turn to a generalized model of cohesion in metals. The total internal energy for a given electronic configuration is determined by a balance between Coulomb ($\propto\frac{1}{a}$) and kinetic ($\propto\frac{1}{a^2}$) energy terms, leading to an energy curve of the form\cite{ashcroft1976}
\begin{equation}\label{cohesion}
E_i(a)=
a_0-\frac{a_1}{a}+\frac{a_2}{a^2},
\end{equation}
where $a$ is the lattice spacing and $a_0$, $a_1$ and $a_2$ are constants.  Here, each $E_i(a)$ curve corresponds to different number of $5f$-electrons, $n_f=0,1,2\dots$, confined to the atomic core.\cite{eriksson1999,svane2007} We can then proceed to obtain the functional form for each $E_i(\nu)$ curve in face-centered cubic $\delta$-Pu (see Appendix) by the substitution of $a=\sqrt[3]{4V}$ and $V=V_0~[1+\nu]$ into Equation~(\ref{cohesion}), from which we obtain
\begin{equation}\label{cohesionvolume}
E_i(\nu)=E_{i,0}+\frac{9K_{i,0}}{2N}\big[1-2[1+\nu-\nu_i]^{-\frac{1}{3}}+[1+\nu-\nu_i]^{-\frac{2}{3}}\big].
\end{equation}\\
The parabolic approximation in Equation~(\ref{cohesionsmall}) is obtained by making a Taylor series expansion of Equation~(\ref{cohesionvolume}) about $\nu-\nu_i$. The energy minimum for each curve is given by $E_{i,0}=a_0-\frac{1}{4}\big[\frac{a^2_1}{a_2}\big]$ while the bulk modulus at the minimum is given by $K_{i,0}=\frac{N}{18}\big[\frac{a^2_1}{a_2}\big]$. Differentiation of Equation~(\ref{cohesionvolume}) yields 
\begin{equation}\label{derivative}
\frac{\partial E_i(\nu)}{\partial\nu}=\frac{3K_{i,0}}{N}\big[[1+\nu-\nu_i]^{-\frac{4}{3}}-[1+\nu-\nu_i]^{-\frac{5}{3}}\big],
\end{equation}\\
which can be considered as a (negative) partial pressure, while further differentiation yields
\begin{equation}\label{secondderivative}
\frac{\partial^2E_i(\nu)}{\partial\nu^2}=\frac{K_{i,0}}{N}\big[5[1+\nu-\nu_i]^{-\frac{8}{3}}-4[1+\nu-\nu_i]^{-\frac{7}{3}}\big].
\end{equation}\\
For $\nu=0$, we obtain the relation
\begin{equation}\label{ambientmoduli}
K_i=K_{i,0}\big[5[1-\nu_i]^{-\frac{8}{3}}-4[1-\nu_i]^{-\frac{7}{3}}\big].
\end{equation}
Theoretical values of $K_i(\nu=0)$ are listed in Table~\ref{table2}.

\begin{table}[ht]
\centering
\begin{tabular}{c c c c c}\\
\hline\hline
$n_{5f}$ & ~~~~$K_i$ (GPa), Ref.\cite{eriksson1999} & ~~~~$K_i$ (GPa), Ref.\cite{svane2007}\\ [0.5ex] 
\hline
0&-&26.0\\
1&-&28.0\\
2&21.8&31.0\\
3&34.5&24.4\\
4&36.4&20.6\\
5&28.5&33.9\\
6&-&25.6\\[1ex]
\hline
\end{tabular}
\caption{{\bf Theoretical bulk moduli parameters}. The theoretical contributions $K_i$ to the bulk modulus at $\nu=0$ for each of the electronic configurations calculated by Eriksson {\it et al.}\cite{eriksson1999} and Svane {\it et al.}.\cite{svane2007} Here, $n_{5f}$ refers to the number of $5f$-electrons artificially confined to the atomic core for each configuration.\cite{eriksson1999,svane2007} Surprisingly, $K_i$ is not found to depend significantly on the calculation method or on $n_{5f}$. Their average is $\bar{K}_i(\nu=0)\approx$~28.2~GPa while their standard deviation is $\sigma K_i=$~5.0~GPa (i.e. $\approx\frac{1}{6}^{\rm th}$ of $\bar{K}_i$).
}
\vspace{2cm}
\label{table2}
\end{table}

The calculation of $K^\prime=NK_i\frac{\partial^3E_i}{\partial\nu^3}$ from the derivative of Equation~(\ref{secondderivative}) yields $K^\prime=4$ at $\nu=0$ for a normal metal. However, the absence of a discernible dependence of the Debye temperature on $x$\cite{harrison2019} suggests that $K^\prime$ is actually closer to zero for the ground state configuration of $\delta$-Pu, thereby further justifying the use of a parabolic approximation for small $\nu$ in the present study. A more accurate experimental determination of $K^\prime$ for the ground state configuration would require low temperature bulk modulus measurements to be performed under pressure.

\subsection{Thermodynamics of multiple configurations}


For the thermal expansion, one differentiates the free energy once with respect to $\nu$ to obtain
\begin{equation}\label{thermalexpansionderivative}
\frac{\partial F_{\rm el}}{\partial\nu}\bigg|_T=-P=\sum_ip_i(\nu)K_i~[\nu-\nu_i],
\end{equation}
where we have again made use of the parabolic approximation given by Equations~(\ref{cohesionsmall}) and (\ref{effectivepressure}). Since the total pressure $P\approx0$ during experiments under ambient conditions, $\nu$ and $\nu_i$ can be separated in Equation~(\ref{thermalexpansionderivative}) to yield
\begin{equation}\label{thermalexpansion}
\nu=\frac{\sum_ip_i(\nu)P_i}{\sum_ip_i(\nu)K_i}.
\end{equation}
Here, the numerator is equivalent to a summation over partial pressures, where the total pressure is ambient pressure. Meanwhile, the denominator is equivalent to the first term of Equation~(\ref{bulkmodulusequation}), meaning that it is equivalent to the bulk modulus that one obtains on neglecting excitations. If we constrain the bulk moduli to be similar for all relevant configurations (i.e. $K_i=K_0$), then the denominator becomes $K_0$ and Equation~(\ref{thermalexpansion}) acquires the much simpler form:
\begin{equation}\label{thermalexpansionsimple}
\nu\approx\sum_ip_i(\nu)\nu_i. 
\end{equation}
It is instructive to express Equation~(\ref{bulkmodulusequation}) in a similarly reduced form by setting $K_i=K_0$ for all configurations and defining $k_{\rm el}=\frac{K_{\rm el}}{K_0}-1$, whereupon we obtain
\begin{equation}\label{bulkmodulusequationsimple}
k_{\rm el}\approx\frac{K_0}{Nk_{\rm B}T}\Big[-\sum_{i}p_i(\nu)\nu^2_i
+\big[\sum_{i}p_i(\nu)\nu_i\big]^2\Big],
\end{equation}
in which the negative term inside the parenthesis dominates. Gr\"{u}neisen's law, in its original form given by $\Gamma=\frac{\alpha_vV_{\rm m}K_T}{C_v}$ is violated because $\alpha_v$, which is the temperature derivative of Equation~(\ref{thermalexpansionsimple}), is proportional to the sum over $\nu_i$ contributions, whose individual values are both positive or negative, thereby giving rise to sign changes.\cite{harrison2019} By contrast, the bulk modulus softening depends on a summation over $\nu_i^2$ contributions, causing Equation~(\ref{bulkmodulusequationsimple}) always to have the same negative sign. The heat capacity, meanwhile, depends only on the energies $E_i$, which are positive (or zero) and indirectly related to $\nu_i$.

In the case of the pressure derivative, the last term on the right-hand-side of Equation~(\ref{pressurederivative}) is given by
\begin{eqnarray}\label{correction}
\delta=3Nk_{\rm B}T\bigg[\big[\sum_ip_i(\nu)P_i(\nu)\big]\big[\sum_ip_i(\nu)K_i\big]~~~~~~~\nonumber\\
-\big[\sum_ip_i(\nu)P_i(\nu)K_i\big]\bigg],
\end{eqnarray}
which vanishes when $K_i$ are the same for all configurations.


\subsection{Additive electronic and phonon bulk moduli}

Since the isothermal bulk modulus is given by $\frac{\partial^2F}{\partial\nu^2}\big|_T$ where the free energy $F=F_{\rm el}+F_{\rm ph}$ is the summation of electronic and phonon contributions, this same summation should generally carry over to the bulk modulus $K_T=K_{\rm el}+K_{\rm ph}$, with the electronic and phonon contributions therefore acting in parallel. The precise form of the dependence of $F_{\rm ph}$ leading to $K_{\rm ph}$ is still an area of active debate.\cite{digilov2019,ida1969,born1939,varshni1970,rose1984,anderson1989,ida1970}

\begin{figure}[!!!!!!!htbp]
\centering 
\includegraphics*[width=.45\textwidth]{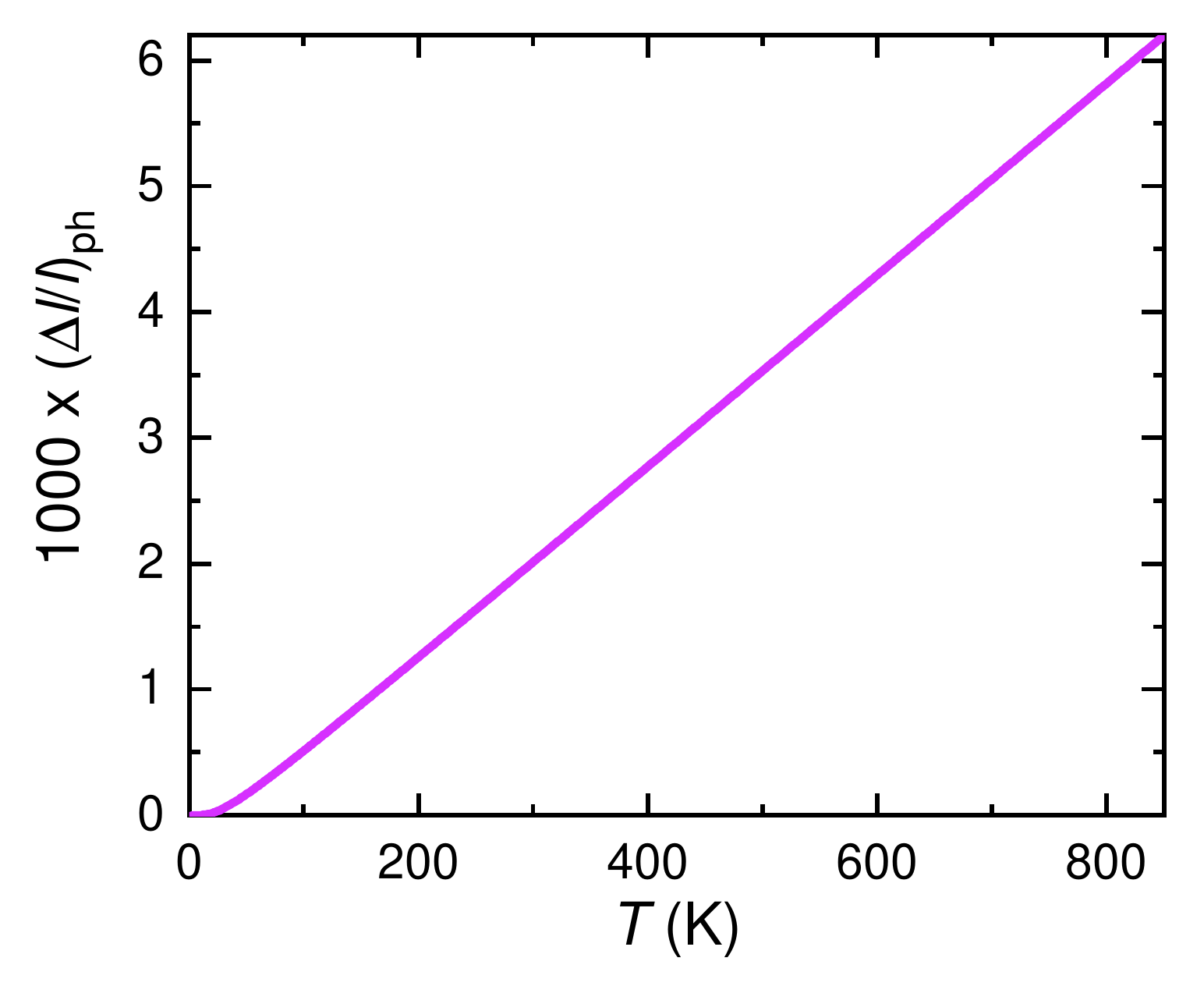}
\caption{The linear thermal expansion of $\delta$-Pu attributable to phonons, determined from a combined fit to thermal expansion and magnetostriction data, and its verification using heat capacity measurements.\cite{harrison2019}}
\label{phononexpansion}
\end{figure}

According to Ida\cite{ida1969}
\begin{equation}\label{phononbulkmodulus}
K_{\rm ph}(T)=K_0\bigg[\frac{T}{T_{\rm 0}}\frac{1}{Q}-1\bigg],
\end{equation}
where $T_{\rm 0}$ is the temperature scale associated with lattice vibrations (see Table~\ref{table1}), $(\Delta l/l)_{\rm ph}$ is the thermal expansion of the lattice attributable to phonons\cite{harrison2019} (see  Fig.~\ref{phononexpansion}) and  $Q$ is the vibrational elongation determined by solving
\begin{equation}\label{Qdetermination}
Q=\frac{T}{T_{\rm 0}}{\rm e}^{2b\big[(\frac{\Delta l}{l})_{\rm ph}+Q\big]}.
\end{equation}

\subsection{Calculating the adiabatic bulk modulus}

\begin{figure}[!!!!!!!htbp]
\centering 
\includegraphics*[width=.45\textwidth]{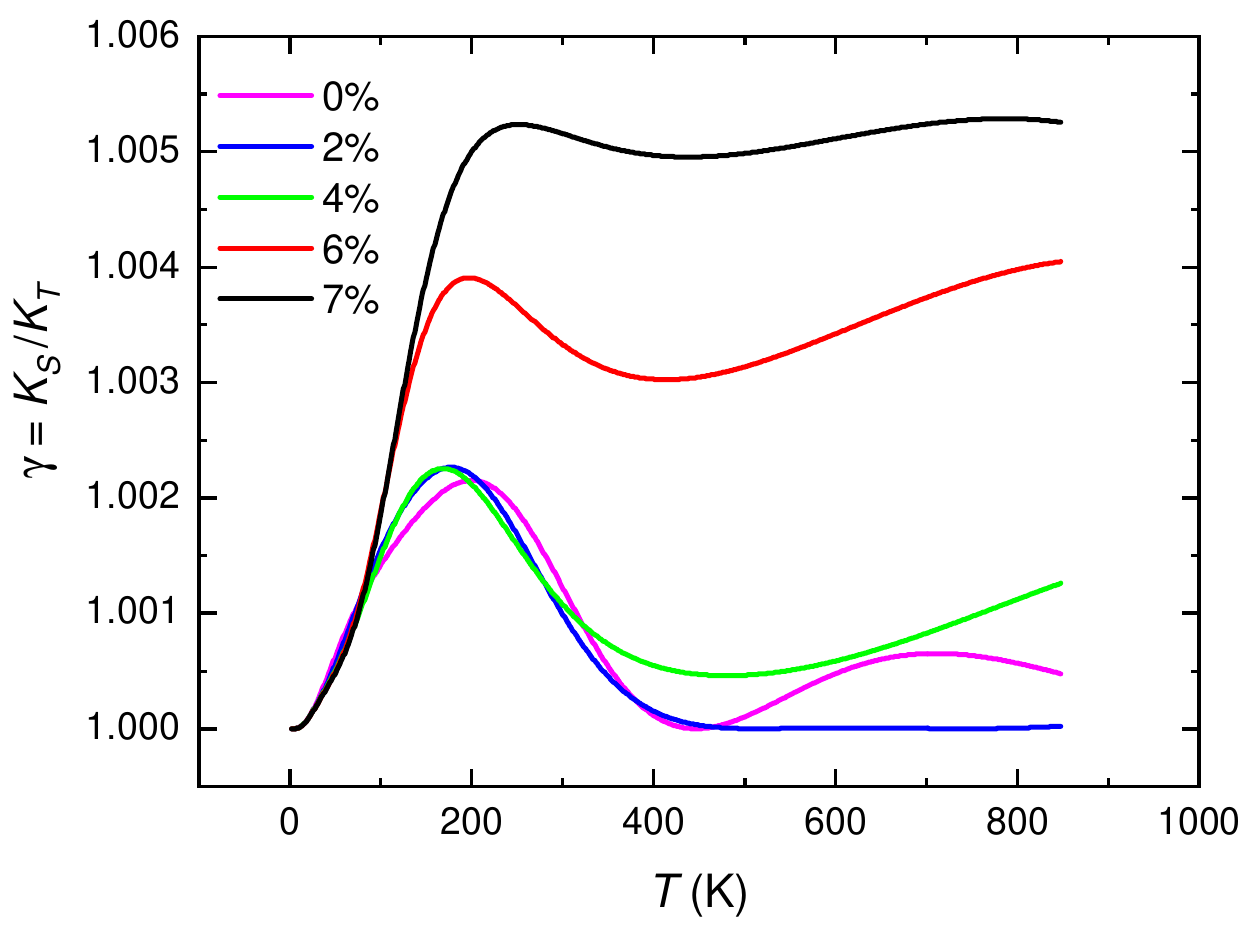}
\caption{Calculated ratio $\gamma=K_S/K_T=C_p/C_T$, according to multiple electronic configurations,\cite{harrison2019} with the concentrations of Ga substituted into $\delta$-Pu indicated in different colors.}
\label{adiabaticcorrection}
\end{figure}

According to basic thermodynamics
\begin{equation}
\gamma=\frac{\alpha_v^2V_{\rm m}TK_T}{C_v}+1
\end{equation}
where $\alpha_v=3\alpha_l$ and $C_v$ values have recently been calculated by Harrison~{\it et al}\cite{harrison2019}, from which we obtain the temperature-dependent $\gamma$ in Fig.~\ref{adiabaticcorrection}. Since $\gamma$ is close to unity, $K_S\approx K_T$. Significantly, for compositions $x\approx$~2\%, the difference between $K_S$ and $K_T$ becomes negligible.

\subsection{Prior models of the bulk modulus softening}

In a prior model of the bulk modulus based on the invar model, the Debye temperature, and consequently the bulk modulus, was assumed to be the probability-weighted sum of the ground state bulk modulus and an excited invar configuration bulk modulus,\cite{lawson2002,lawson2006,lawson2019} which is equivalent to the first term of Equation~(\ref{bulkmodulusequation}). In the absence of any other terms contributing to the bulk modulus, Lawson {\it et al.} were able to approximately account for the experimentally observed softening of the bulk modulus by setting $K_i=0$ for the excited invar configuration. Such a value is expected to occur only at a very large positive volume strain of $\nu=\frac{61}{64}+\nu_i$, which we obtain by equating Equation~(\ref{secondderivative}) to zero. A volume strain this large ($\nu\sim$~100\%) corresponds to an actual volume of order $V\sim$~50~$\AA^3$, which is significantly larger than the value $V_i\approx$~20~$\AA^3$ for the excited configuration of $\delta$-Pu obtained in fitting the invar model to the thermal expansion.\cite{lawson2002,lawson2006}

In another more recent model of the bulk modulus based on the disordered local moment model,\cite{migliori2016} the volume was assumed to remain approximately constant with the different configurations corresponding to a continuum of states with different degrees of orbital compensation of the local moment.\cite{soderlind2008} In this model, the softening of the bulk modulus occurs in response to a reduction in the moment with increasing temperature. However, since the degree of softening in this model is predicted to be reduced for samples with a larger negative contribution to the thermal expansion, which generally occurs for samples with lower concentrations $x$ of Ga,\cite{lawson2002} the predicted Ga-dependent trend is opposite to that found experimentally (plotted in Fig.~\ref{experimentalbulkmod}b).\cite{soderlind2010}

\subsection{Offsets in the measured bulk modulus}

Significant vertical displacements between bulk moduli values for different samples of the same composition together with the absence of a trend in $x$ near room temperature in Fig.~\ref{experimentalbulkmod}a, indicate that extrinsic factors unrelated to $K_{\rm el}$ and $K_{\rm ph}$ contribute random vertical offsets to the experimental data that are of order 1 or 2~GPa. Similar observations have been made in measuring control samples made of aluminum. In Fig.~\ref{experimentalbulkmod}b, the extrinsic and phonon contributions are removed from the analysis of the $x=$~2.36\%, 3.30\% and 4.64\% datasets by subtracting the values of the bulk moduli at $T=$~280~K.

\subsection{Questions regarding pressure-dependent data}

There are uncertainties in how some of the experimental compositions in Fig.~\ref{pressurebulkmod} can be compared with the model. For instance, the $x=$~3.5\% Ga-stabilized sample, for which $K^\prime\sim+4$ is positive in contrast to the other $\delta$-Pu samples,\cite{zhang2019} has a substantially lower volume ($V\approx$~24.2~\AA$^3$) than that ($V\approx$~24.6~\AA$^3$) previously found for samples of nominally the same composition (interpolating between $x=$~2 and 4\%).\cite{lawson2002} Also, the lowest error bar ($K^\prime=$~$-4$~$\pm$~$2$) is obtained for a heavily Am-stabilized $\delta$-Pu sample (likely due to the the wider range of accessible pressures), yet its thermal expansion and elastic properties remain largely unexplored.\cite{hecker2004}\\

\section{Acknowledgements}

The work was performed under the Los Alamos National Laboratory LDRD program: project ``20180025DR.'' Measurements were performed at the National High Magnetic Field Laboratory, which is supported by the National Science Foundation, Florida State and the Department of Energy. N. H. thanks Albert Migliori, Boris Maiorov, Angus Lawson and Paul Tobash for insightful discussions.


%

\end{document}